\documentclass[reprint,
superscriptaddress,
preprintnumbers,
 amsmath,amssymb,
pra,
floatfix,
]{revtex4-2}

\usepackage{graphicx}
\usepackage{dcolumn}
\usepackage{subfigure}
\usepackage{bm}
\usepackage[utf8]{inputenc}  
\usepackage[unicode=true, bookmarks=false, breaklinks=false, pdfborder={0 0 1},colorlinks=false]{hyperref}
\usepackage{adjustbox} 
\usepackage{url}
\usepackage{natbib}
\usepackage{booktabs}
\usepackage{multirow}
\usepackage{float}
\usepackage{amsmath}

\usepackage{xcolor}
\usepackage{soul}
\usepackage{etoolbox}
\usepackage{derivative}
\usepackage{textcomp}

\begin{document}

\preprint{FeMn-ratio}

\title{
Giant magnetocaloric effect in the (Mn,Fe)NiSi-system 
}

\newcommand{\uppsala}{Department of Materials Science and Engineering, Uppsala University, Box 35, SE-751 03 Uppsala, Sweden}

\newcommand{\chem}{Department of Chemistry – Ångström Laboratory, Uppsala University, Box 538,  SE-751 21 Uppsala, Sweden}

\newcommand{\phys}{Department of Physics and Astronomy, Uppsala University, Box 516, SE-751 20 Uppsala, Sweden}

\newcommand{\abo}{Physics, Faculty of Science and Engineering, Åbo Akademi University, FI-20500 Turku, Finland }

\newcommand{\orebro}{School of Science and Technology, \"{O}rebro University, SE-701 82 \"{O}rebro, Sweden}

\newcommand{\serg}{Department of Physics, Chemistry and Biology (IFM),
Link\"{o}ping University, SE-581 83, Link\"{o}ping, Sweden
}

\newcommand{\contribution}{These authors contributed equally to this work.}

\author{Sagar Ghorai}
\email[]{sagar.ghorai@angstrom.uu.se}
\affiliation{\contribution}
\affiliation {\uppsala}

\author{Rafael Martinho Vieira}
\email[]{rafael.vieira@physics.uu.se}{}
\affiliation{\contribution}
\affiliation{\phys}
\affiliation{\abo}

\author{Vitalii Shtender}
\affiliation{\chem}

\author{Erna K. Delczeg-Czirjak}
\affiliation{\phys}

\author{Heike C. Herper}
\affiliation{\phys}

\author{Torbj\"orn Bj\"orkman}
\affiliation{\abo}

\author{Sergei I. Simak}
\affiliation{\phys}
\affiliation{\serg}

\author{Olle Eriksson}
\affiliation{\phys}

\author{Martin Sahlberg}
\affiliation{\chem}

\author{Peter Svedlindh}
\affiliation{\uppsala}

\date{\today}



\maketitle


{\bf The search for energy-efficient and environmentally friendly cooling technologies is a key driver for the development of magnetic refrigeration based on the magnetocaloric effect (MCE). This phenomenon arises from the interplay between magnetic and lattice degrees of freedom that is strong in certain materials, leading to a change in temperature upon application or removal of a magnetic field. Here we report on a new material, Mn$_{1-x}$Fe$_x$NiSi$_{0.95}$Al$_{0.05}$, with an exceptionally large isothermal entropy at room temperature. By combining experimental and theoretical methods we outline the microscopic mechanism behind the large MCE in this material. It is demonstrated that the competition between the Ni$_2$In-type hexagonal phase and the MnNiSi-type orthorhombic phase, that coexist in this system, combined with the distinctly different magnetic properties of these phases, is a key
parameter for the functionality of this material for magnetic cooling. 
}

Materials exhibiting a large magnetic field-induced isothermal entropy change ($\Delta S_M$) are classified as giant magnetocaloric effect (GMCE) materials and are desired for magnetic refrigeration near room temperature  \cite{gutfleisch2011magnetic}. One of the main factors contributing to the GMCE is a magneto-structural transition (MST), where a structural phase transition coincides with a magnetic phase transition, resulting in a large change in the magnetization and magnetic entropy. However, to achieve a GMCE, the system must meet two requirements: a large difference in the magnetic properties of the two structural phases and a narrow temperature range between the magnetic ordering temperature ($T_C$) and the structural transition temperature ($T_{st}$). These conditions most likely arise in systems with a strong coupling between the magnetic and lattice degrees of freedom. 

\begin{figure}
    \centering
    \begin{adjustbox}{max width=\textwidth, max height=0.8\textheight}
        \includegraphics{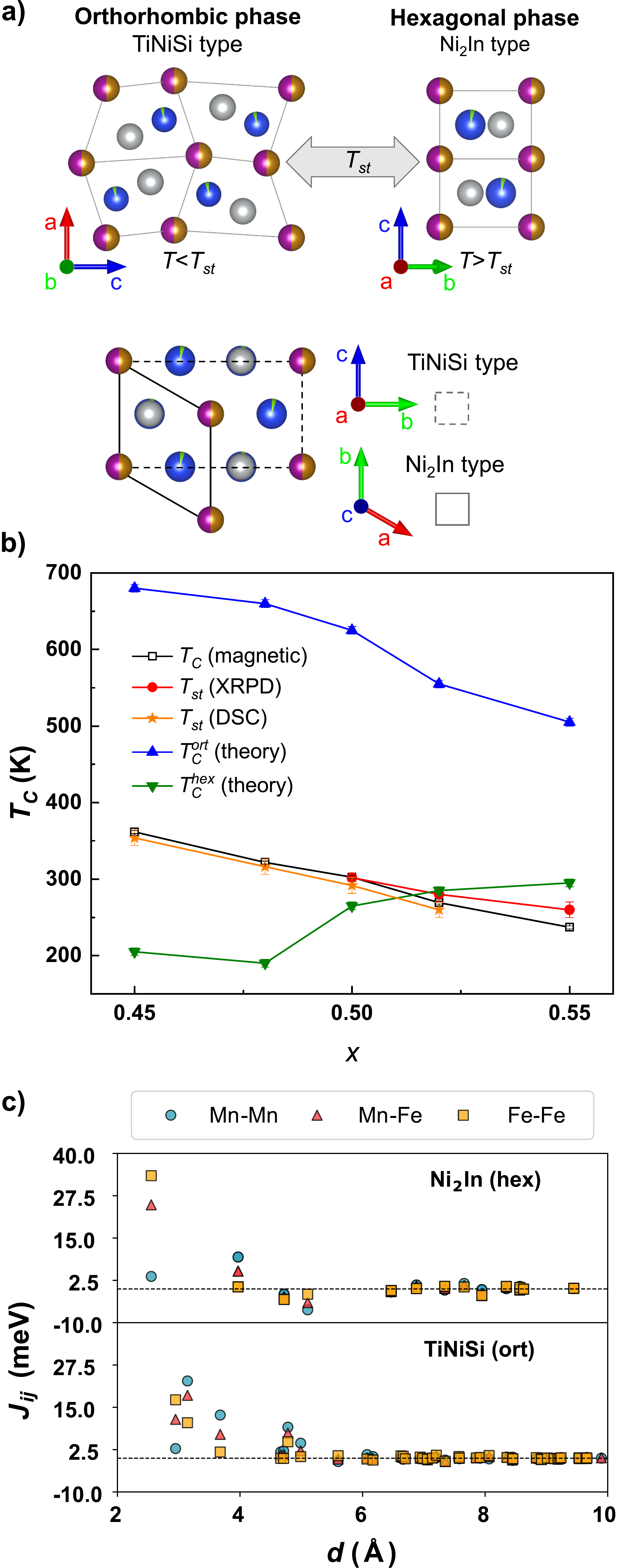}
    \end{adjustbox}

    \caption{(a) Hexagonal and orthorhombic phases of Mn$_{0.5}$Fe$_{0.5}$NiSi$_{0.95}$Al$_{0.05}$  obtained from XRPD refinement. Top panel highlights the alterations in the arrangement of the atoms due to the phase transition. Bottom panel illustrates the relation between unit cells. The colours represent the site occupation by the different chemical elements: Mn (pink), Fe (orange), Ni (grey), Si (blue), and Al (green). Plotted with VESTA \cite{momma2011vesta}. (b) Calculated and experimental transition temperatures versus the amount of Fe ($x$) for the Mn$_{1-x}$Fe$_x$NiSi$_{0.95}$Al$_{0.05}$ system. (c) Calculated exchange parameters $J_{ij}$ for  Mn$_{0.5}$Fe$_{0.5}$NiSi$_{0.95}$Al$_{0.05}$ in the hexagonal (top) and the orthorhombic (bottom) phases.}
    
    \label{Theory}
    
\end{figure}

The Mn$TX$-system ($T=$ Co, Ni and $X=$ Ge, Si) represents a promising class of intermetallic shape-memory compounds that may exhibit an MST. The parent compound, MnNiSi, is known to undergo a structural phase transition from a low-temperature orthorhombic phase (TiNiSi-type, space group $Pnma$) to a high-temperature hexagonal phase (Ni$_2$In-type, space group $P63/mmc$). The orthorhombic phase can be viewed as a distorted version of the hexagonal phase, with the lattice parameters related as: $a_{orth}\equiv c_{hex}$, $b_{orth}\equiv a_{hex}$, and $c_{orth}\equiv \sqrt{3} a_{hex}$ \cite{liu2016realization}. The two phases are illustrated in Fig.~\ref{Theory}(a), highlighting similarities and differences between the structures.
In the parent compound, the structural, $T_{st} = 1206$ K, and magnetic transition, $T_C = 622$ K, temperatures differ greatly \cite{johnson1975diffusionless,johnson1973magnetic}, indicating that the expected coupled MST is not present.


Recently, it has been found that both $T_C$ and $T_{st}$ can be tuned by Mn-site \cite{samanta2013mn1}, Ni-site \cite{zhang2008magnetostructural,liu2012stable} and Si-site \cite{biswas2019designed} chemical substitutions in the MnNiSi parent compound. In the Mn$_{0.5}$Fe$_{0.5}$NiSi$_{1-x}$Al$_x$ compound, an introduction of $1.5 - 2.3 \%$ Al shifts both $T_C$ and $T_{st}$ to near room temperature \cite{biswas2019designed} and a MST becomes possible. Owing to the evolution of the orthorhombic and hexagonal phases with temperature and their coexistence in the magnetic transition region, the description of $T_C$ for the composite phase (where the two structures coexist) is complicated and has so far not been clearly described. 
The coexistence of the orthorhombic and hexagonal phases near the magnetic transition indicates that the magnetic exchange interactions ($J_{ij}$) of the composite system will contain contributions from  both phases. Moreover, since the relative weight-$\%$ of the two structural phases varies with temperature in the magnetic transition region, a concomitant temperature dependent variation of $J_{ij}$ may be possible, which together with the relative orientation of the magnetic moments will decide $T_C$ of the composite phase, potentially leading also to a significant MCE.

The present work identifies the Mn$_{1-x}$Fe$_x$NiSi$_{0.95}$Al$_{0.05}$ (MnFeNiSiAl) system as a potential GMCE compound near room temperature. Using a combination of first-principles calculations and Monte Carlo simulations, the expected variation of $T_C$ in the composite phase approach has been calculated and verified experimentally. A large value of the isothermal entropy change is found, and explained by the sharpness of the MST. Additionally, by comparing the total energy from the electronic structure calculations, the atomic site occupancy of the Fe/Mn and Ni atoms in the MnFeNiSiAl system has been determined (\textcolor{green}{details in Supplementary Information (SI) \ref{SI:site_occ}}).


\subsection*{First-principles calculations}

The unit cell and atom projected magnetic moments have been calculated by ab-initio theory, both for the orthorhombic and hexagonal phases, and are shown in Table~\ref{tab:summary_X}. A weak dependence between the magnetic moment and the Fe concentration is observed, with the total magnetic moment decreasing  with increasing amount of Fe, which is in accordance with experimental results for the orthorhombic phase reported here and in Ref.~\cite{Vaez2013}. 

Contrary to what is often observed in GMCE materials, the magnetic moments for Mn and Fe sites are similar in the orthorhombic and hexagonal phases. This is in agreement with the results for the MnNiSi and FeNiSi systems in Ref.~\cite{Fortunato2023}, but in contrast with the results reported for Mn$_{0.5}$Fe$_{0.5}$NiSi system in Ref.~\cite{biswas2021controlling}.

Besides variations in the local magnetic moments, a significant difference in the magnetization between phases can be achieved if the phases exhibit a difference in the type of magnetic order or if they have different magnetic ordering  temperatures. This is determined by the interatomic exchange parameters of a Heisenberg Hamiltonian, and since ab-initio theory has the capability to evaluate these parameters\cite{szilva2022quantitative}, it is possible to obtain a parameter-free spin Hamiltonian that, when coupled to Monte Carlo (MC) simulations, give reliable information of finite temperature magnetism \cite{binder1993monte}. 


\begin{table*}[ht]
\centering
\caption{Summary of calculated magnetic properties and  geometrical parameters from XRPD refinement for different amounts of Fe $x$ in the Mn$_{1-x}$Fe$_x$NiSi$_{0.95}$Al$_{0.05}$ system.}
\label{tab:summary_X}

\begin{ruledtabular}
\begin{tabular}{cccccccccccccccc}
 & \multicolumn{8}{c}{\textbf{Orthorhombic phase}} & \multicolumn{7}{c}{\textbf{Hexagonal phase}}\\
 \cline{2-16}
$x$ & $a$ & $b$ & $c$ & $m_{Mn}$ & $m_{Fe}$ & $m_{Ni}$ & $m_{tot}$ & \multicolumn{1}{c|}{$T_{C}$} & $a$ & $c$ & $m_{Mn}$ & $m_{Fe}$ & $m_{Ni}$ & $m_{tot}$ & $T_{C}$ \\
   & (\AA) & (\AA) & (\AA) & ($\mu_B$) & ($\mu_B$) & ($\mu_B$) & ($\mu_B$/f.u.) & \multicolumn{1}{c|}{(K)} & (\AA) & (\AA) & ($\mu_B$) & ($\mu_B$) & ($\mu_B$) & ($\mu_B$/f.u.) & (K) \\ \hline
0.45 &  5.656 & 3.659 & 6.929 & 2.514 & 1.889 & 0.158 & 2.282 & \multicolumn{1}{c|}{680$\pm$ 5} & 3.978 & 5.114 & 2.450 & 1.901 & 0.109 & 2.220 & 205$\pm$ 5 \\
0.48 & 5.627 & 3.667 & 6.934 & 2.506 & 1.873 & 0.154 & 2.247 & \multicolumn{1}{c|}{660$\pm$ 5} & 3.976 & 5.106 & 2.436 & 1.880 & 0.106 & 2.180 & 190$\pm$ 5 \\
0.50 &  5.598 & 3.675 & 6.940 & 2.462 & 1.824 & 0.164 & 2.201 & \multicolumn{1}{c|}{625$\pm$ 5} & 3.972 & 5.106 & 2.432 & 1.866 & 0.104 & 2.159 & 265$\pm$ 5 \\
0.52 & 5.588 & 3.679 & 6.907 & 2.387 & 1.771 & 0.211 & 2.172 & \multicolumn{1}{c|}{555$\pm$ 5} & 3.970 & 5.097 & 2.410 & 1.841 & 0.101 & 2.121 & 285$\pm$ 5 \\
0.55 & 5.495 & 3.771 & 6.856 & 2.495 & 1.872 & 0.131 & 2.178 & \multicolumn{1}{c|}{505$\pm$ 5} & 3.970 & 5.093 & 2.408 & 1.839 & 0.100 & 2.101 & 295$\pm$ 5 \\ 
\end{tabular}
\end{ruledtabular}
\end{table*}

From a comparison of the calculated exchange parameters, shown in Fig.~\ref{Theory}(c), it is observed that in contrast to the orthorhombic phase, the hexagonal structure possesses non-negligible antiferromagnetic couplings which explains the low-temperature non-collinear spin structure of this phase, being a consequence of competition between the first (in-plane) and second (between hexagonal planes) nearest neighbours with the third and fourth nearest neighbours (\textcolor{green}{see SI.\ref{SI:spins_dist} for details}).
Moreover, it is observed that for the hexagonal phase the exchange parameters are much stronger than for the orthorhombic phase. In addition, the long range nature of the exchange interactions is important for this system; 
including only exchange parameters within a radius of 6~\AA~ in the Monte Carlo simulations (see below) decreases the magnetic ordering temperature to around 120$\pm$20 K. 

From Monte Carlo simulations, using the calculated atomic moments and the exchange parameters of Fig.~\ref{Theory}(c), an overall decrease of the calculated $T_C$ with increasing amount of Fe is observed for the orthorhombic phase, cf. Fig.~\ref{Theory}(b).  This trend agrees with the experimental results (see Fig.~\ref{Theory}(b)), although the calculated ordering temperature is much too large compared to the observations, and the decrease of $T_C$ as function of Fe concentration is smaller than the one measured. Note that the temperature range for the calculated $T_C$ of the orthorhombic phase (see also Table \ref{tab:summary_X}) is close to the value of the orthorhombic parent compound, MnNiSi ($\approx$ 600 K). The calculated magnetic transition temperature for the hexagonal phase shows a trend which is opposite to that of the orthorhombic phase, with an increase with increasing amount of Fe. The ordering temperature for this phase is also seen to be much lower compared to the data of the orthorhombic structure, hence displaying a surprisingly strong phase dependence on the ordering temperature. 

\subsection*{Structural and magnetic phase diagram}

\begin{figure*}[ht]
    \centering
    \includegraphics[width=\linewidth]{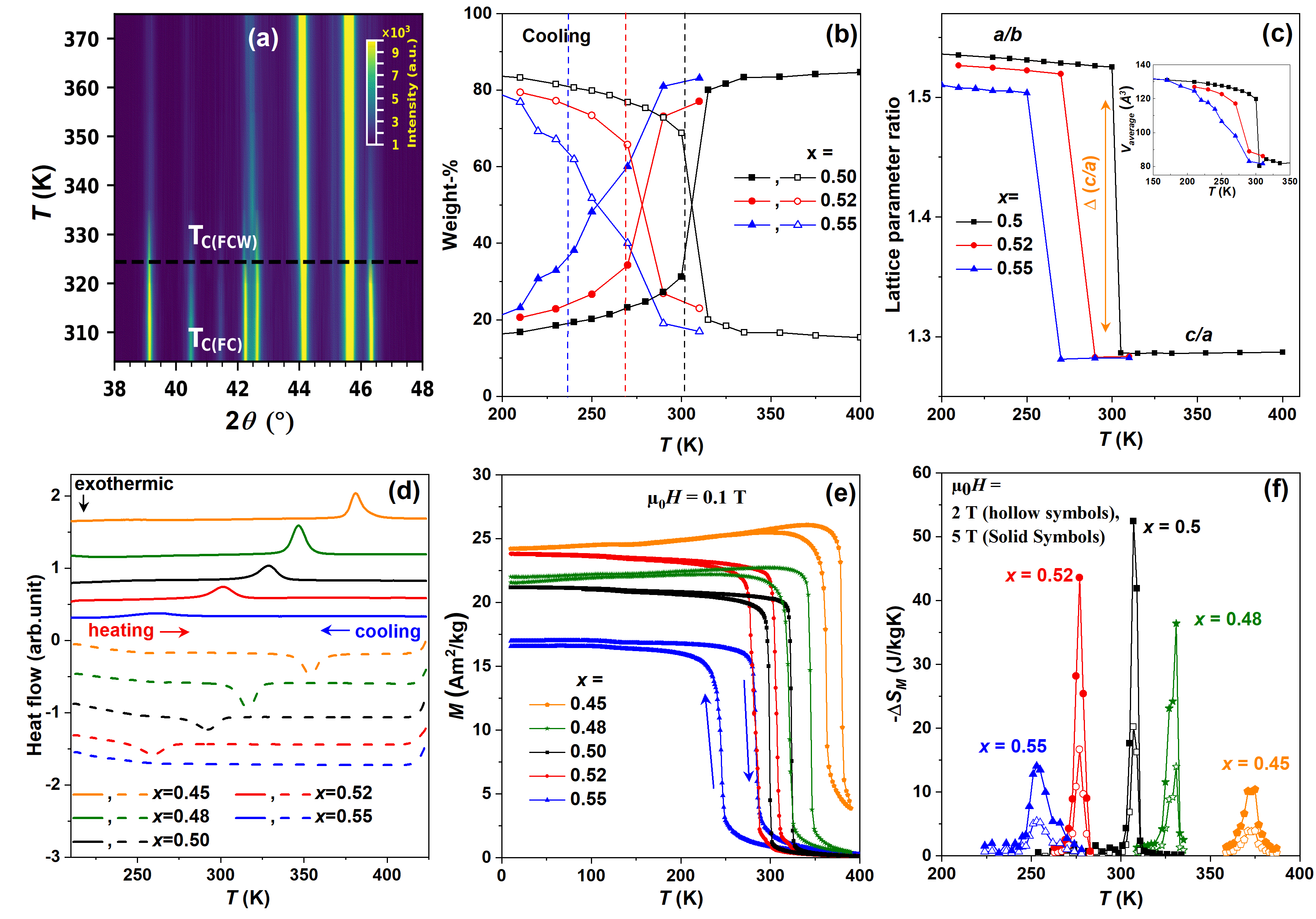}
    \caption{(a) Temperature dependent X-ray powder diffraction pattern of Mn$_{0.5}$Fe$_{0.5}$NiSi$_{0.95}$Al$_{0.05}$. The dashed horizontal line indicates $T_C$ during heating ($T_{C(FCW)}$); $T_C$ during cooling (FC) is at $300$ K. (b) temperature dependent weight-\% of the two structural phases for the $x= 0.5, 0.52$ and $0.55$ compounds calculated from the XRPD data recorded during the cooling cycle. The hollow and solid symbols represent the orthorhombic and hexagonal phases, respectively. The dashed  color-coded vertical lines represent $T_C$ of the corresponding compounds. (c) temperature dependent variation of the hexagonal lattice parameter ratio $c/a$ and the  equivalent lattice parameter ratio of the orthorhombic phase $a/b$. The inset shows the temperature dependence of the average volume of the unit cell for the different compounds. The error in the phase weight-\% is less than $0.5\%$ and the error in the lattice parameter data is in the order of $10^{-3}$ Å, therefore not included in the figure. (d) Differential scanning calorimetry versus temperature for the studied compounds during cooling and heating cycles. (e) Temperature dependent magnetization and (f) temperature dependent isothermal entropy change for the different compounds.} 
    \label{XRD_MCE} 
\end{figure*}

To experimentally determine the crystallographic phase contribution, temperature dependent X-ray powder diffraction (XRPD) measurements have been performed on all compounds, that were found to exhibit two major crystallographic phases; the TiNiSi-type orthorhombic phase and the Ni$_2$In-type hexagonal phase. However, a small amount of MnZn$_2$-type hexagonal impurity phase has also been observed, which \textcolor{green}{is discussed in SI.\ref{SI:impurity}}. The relative amount of the two major phases is found to be distinctly temperature dependent. Moreover, the $x= 0.5$ compound is found to undergo a structural phase transition from the low-temperature orthorhombic to the high-temperature hexagonal phase as shown in Fig. \ref{XRD_MCE}(a) . 

Figure \ref{XRD_MCE}(b) shows the temperature dependence of the weight-\% of the two structural phases, where the crossing of the curves for the different compounds 
indicates $T_{st}$ for the corresponding compound. $T_{st}$ decreases with increasing amount of Fe. The structural phase change is driven by thermal energy, enabling the distorted orthorhombic structure to transform into the hexagonal structure. A lower value of the $c_{hex}/a_{hex}$ ratio indicates a more stable hexagonal structure \cite{samanta2015hydrostatic}. From Fig. \ref{XRD_MCE}(c), a decreasing trend of $c_{hex}/a_{hex}$ is observed with increasing amount of Fe. The difference between the $c_{hex}/a_{hex}$ ratio in the hexagonal phase and the $a_{orth}/b_{orth}$ ratio in the orthorhombic phase is proportional to the thermal energy required for the structural phase change. Therefore, the decrease of the difference between $c_{hex}/a_{hex}$ and $a_{orth}/b_{orth}$, denoted as $\Delta (c/a)$ ($\Delta(c/a)= a_{orth}/b_{orth}-c_{hex}/a_{hex}$), see the orange arrow in Fig. \ref{XRD_MCE}(c)), with increasing Fe amount indicates a similar decrease of the thermal energy required for the structural phase change to occur. This also explains the decrease of $T_{st}$ with increasing amount of Fe as seen in Fig. \ref{XRD_MCE}(c) and Fig. \ref{Theory}(b).

The decrease of the hexagonal lattice parameter ratio $c/a$ with increasing Fe amount leads to a decrease of the interlayer distance for the Mn/Fe atoms separated by a Ni-Si layer along the $ab$-plane (cf. Fig. \ref{Theory}(a)). Since the magnetic properties of the MnNiSi system are highly sensitive to the Mn-Mn distance 
\cite{barcza2010giant}, the structural phase change will influence the magnetic exchange parameters of the system. In Fig. \ref{XRD_MCE}(e), the temperature dependent magnetization for the studied compounds is shown. The temperature hysteresis of the magnetic transition, while decreasing and increasing the temperature,  indicates a first-order nature of the phase transition.  Also, $T_C$ decreases with increasing amount of Fe, similar to the trend of $T_{st}$. Below we discuss how the structural transition influences the magnetic transition at $T_C$.

Significantly, as revealed in Fig. \ref{XRD_MCE}(b) a non-negligible fraction ($\approx$ 20\%) of the hexagonal phase remains at low temperature in all of the studied compounds. A similar observation was reported in Ref.~\cite{Zhang2014} for the Mn$_{0.5}$Fe$_{0.5}$NiSi$_{1-x}$Al$_x$ system, where the two phases were shown coexist at room temperature for Al  concentrations between $0.045$ and $0.055$. Below and above this concentration interval, the orthorhombic and hexagonal phases dominated, respectively, highlighting the role of Al for the stability of the hexagonal phase. Interestingly, it was observed in the same work that $T_C$ decreases with increasing amount of Al ~\cite{Zhang2014}. Since neither Al nor Si contributes directly to the magnetic properties and since the low level of doping considered did not induce any significant  change of the lattice parameters, it is reasonable to assume that the magnetic properties of the pure orthorhombic and hexagonal phases did not change to any significant extent. This hints that the interaction between the two phases plays a significant role in the magnetic properties of the composite compound.  

In addition to XRPD, the results from temperature dependent differential scanning calorimetry (DSC) of the studied compounds are shown in Fig. \ref{XRD_MCE}(d). As there is no magnetic field applied during the DSC measurements, the peaks observed in Fig. \ref{XRD_MCE}(d) are mostly indicative of  structural phase transitions. As seen in Fig. \ref{XRD_MCE}(e), the temperature hysteresis of the structural transition yields a corresponding temperature hysteresis of the magnetic transition.
Moreover, as the structural phase transition temperature $T_{st}$ of the compounds occurs in close vicinity of $T_C$ (cf. Fig. \ref{XRD_MCE}(b)), a coupled MST can be expected in the MnFeNiSiAl compounds. The strength of the MST mostly depends on how close $T_C$ and $T_{st}$ are, along with how sharp the transitions are with respect to temperature. As seen in Fig. \ref{XRD_MCE}(b), the sharpest structural transition is observed for the $x=0.5$ compound, which along with the smallest separation between $T_C$ and $T_{st}$ imply that this compound exhibits the strongest MST. The $x=0.55$ compound in comparison exhibits a broad structural transition, which together with more separated values for $T_C$ and $T_{st}$ imply a  weaker MST. Therefore, it is expected that the $x=0.5$ compound will exhibit the largest GMCE among the studied compounds.

\subsection*{GMCE}

The magnetocaloric effect can be described as a system-dependent spin-lattice  interaction phenomenon. Under adiabatic conditions, the sum of the magnetic and lattice entropies of a system is conserved (here neglecting the electronic entropy contribution). Therefore, a change in magnetic entropy with the application or removal of a magnetic field will induce a change in the lattice entropy and hence the temperature of the material. This change in temperature is related to the change of the isothermal entropy ($\Delta S_M$) of the system,  which is an important characteristic of the magnetocaloric effect. The magnetic field induced $\Delta S_M$ can be derived from the Maxwell relation,\cite{ghorai2022direct}

\begin{equation}
\label{eq.1}
    \Delta S_M(T, H_f) = - \mu_0\int_{0}^{H_f} \pdv{M}{T}_H \,dH .
\end{equation}.

\begin{table}[ht]
\caption{Magnetocaloric properties of the here studied compounds (*) compared with data reported for other GMCE materials near room temperature, at an external magnetic field of  $\mu_{0}H = 2$~T.}
\vspace{3pt}
\centering

\begin{ruledtabular}
\begin{tabular}{ccccc}

 Sample&   $T_{C}^{FC}$&  $-\Delta S_{M}$ & $RCP$ & Ref.\\
&(K) & (J/kgK)& (J/kg)& \\
\hline

Mn$_{0.55}$Fe$_{0.45}$NiSi$_{0.95}$Al$_{0.05}$&    362&       3.93&   36.9&    *\\
Mn$_{0.52}$Fe$_{0.48}$NiSi$_{0.95}$Al$_{0.05}$&    322&       14.0&   82.9&    *\\
Mn$_{0.50}$Fe$_{0.50}$NiSi$_{0.95}$Al$_{0.05}$&    300&       20.2&   86.9&    *\\
Mn$_{0.48}$Fe$_{0.52}$NiSi$_{0.95}$Al$_{0.05}$&    269&       16.7&   85.2&    *\\
Mn$_{0.45}$Fe$_{0.55}$NiSi$_{0.95}$Al$_{0.05}$&    237&       5.4&   64.5&    *\\
Mn$_{0.50}$Fe$_{0.50}$NiSi$_{0.95}$Al$_{0.05}$&    $\sim$316&       $\sim$16&   $\sim$70&    \cite{biswas2019designed}\\
Mn$_{0.5}$Fe$_{0.5}$NiSi$_{0.94}$Al$_{0.06}$B$_{0.005}$&    $\sim$280&       22&   66&    \cite{biswas2021controlling}\\
Mn$_{0.6}$Fe$_{0.4}$NiSi$_{0.93}$Al$_{0.07}$&    $\sim$245&       9.3&   -&    \cite{ghosh2020magnetostructural}\\
Fe$_{0.95}$V$_{0.05}$MnP$_{0.5}$Si$_{0.5}$&    318&       9.1&   103&    \cite{PhysRevB.107.104409}\\
FeMn$_{0.95}$V$_{0.05}$P$_{0.5}$Si$_{0.5}$&    322&       13.1&   130&    \cite{PhysRevB.107.104409}\\
Gd&      295&        6.1&   240&    \cite{gschneidner2000magnetocaloric}\\
La(Fe$_{0.98}$Mn$_{0.02}$)$_{11.7}$Si$_{1.3}$H&      312&        13&   -&    \cite{shen2009recent}\\
Fe$_{80}$Pt$_{20}$&      290&        $\sim 10$&   -&    \cite{rong2007temperature}\\
Mn$_{1.2}$Fe$_{0.8}$P$_{0.75}$Ge$_{0.25}$&      288&        20&   -&    \cite{trung2009tunable}\\
MnFeP$_{0.52}$Si$_{0.48}$&      268&        10&   -&    \cite{cam2008structure}\\
La$_{0.5}$Pr$_{0.2}$Ca$_{0.1}$Sr$_{0.2}$MnO$_3$&    296&       1.8&   147&    \cite{skini2020large}\\

\end{tabular}
\end{ruledtabular}

\label{MCEtable}
\end{table}

From Eq. (\ref{eq.1}), it is clear that a high value of $\Delta S_M$ requires a large change of the temperature and magnetic field-dependent magnetization. As described in the previous section, the strongest MST is expected for the $x=0.5$ compound, and the largest value of $\Delta S_M$ is also observed for this compound (cf. Fig. \ref{XRD_MCE}(f)). The MST is also associated with a change in lattice volume. The inset of Fig. \ref{XRD_MCE}(c) shows the change in average lattice volume (scaled with the weight-\% for each phase) for the different compounds. Noticeably, the lower and upper limits of the lattice volume for the compounds are almost the same owing to the small difference in chemical composition between the compounds. However, the volume change is sharper for the $x = 0.5$ compound compared to the other compounds. All these facts indicate  that the strongest MST is found for the $x = 0.5$ compound. As a result, the highest value of  $\Delta S_M$ is observed (cf. Fig. \ref{XRD_MCE}(f)) for the $x=0.5$ compound, followed by the $x = 0.52$ and $x = 0.55$ compounds in ascending order. A similar trend of $\Delta S_M$ is observed for compounds where the Fe content is lower than $x = 0.5$, i.e. this trend is observed for the $x = 0.48$ and $x = 0.45$ compounds also. This suggests that below and above the value of $x=0.5$, the structural transition becomes less and less sharp, leading to lower values of $\Delta S_M$.

The relative cooling power ($RCP$), defined as  \cite{ghorai2020field},
\begin{equation}
    RCP = -\Delta S_{M}^{max} \times \Delta T_{FWHM},
\end{equation}
can be used to quantify the temperature range over which the GMCE is useful for magnetic refrigeration. Table \ref{MCEtable} lists values of $\Delta S_M$ and $RCP$ for the here studied compounds, using Eq. (1) together with experimental isothermal magnetisation measurements. The experimental data of the FeMnNiSiAl system is in Table \ref{MCEtable} compared to data reported for other GMCE materials. To the best of our knowledge, compared to other GMCE compounds, the $x=0.5$ compound exhibits one of the largest  values of $\Delta S_M$ ever reported, and the relative cooling power is competitive to that of other systems listed in Table \ref{MCEtable}.

\subsection*{Coupled phase transition}

\begin{figure}[!h]
    \centering
    \includegraphics[width=\linewidth]{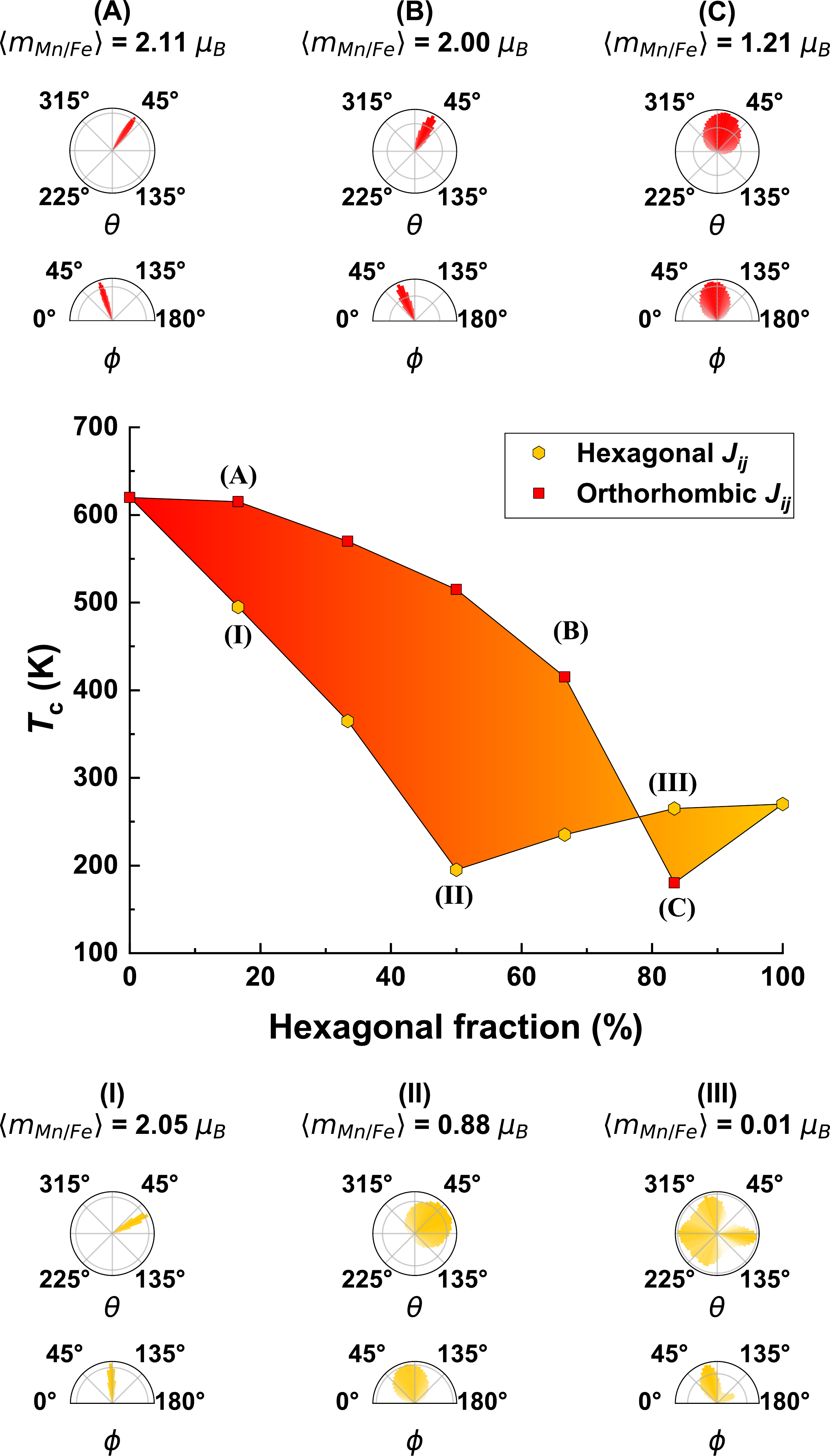}
    \caption{Calculated magnetic transition temperature from  Monte Carlo simulations plotted against the amount of hexagonal phase in the composite. As the interface exchange parameters $J_{ij}^{inter}$ between the hexagonal and orthorhombic phases are unknown, two extreme cases are considered. In the middle panel, the orange line represents when $J_{ij}^{inter} = J_{ij}^{ort}$ and the blue line represents when $J_{ij}^{inter} = J_{ij}^{hex}$. These cases establish a temperature range where the true value should be. In the upper (bottom) panel; A, B, and C (I, II, III) depict the statistical spin distribution in spherical coordinates of the thermalized magnetic spin configuration at $25$ K for the interface $J_{ij}^{inter}$ set of the orthorhombic (hexagonal) phase. The respective average magnetization per magnetic site is also shown.}
    \label{fig:mixing_TC}
\end{figure}

\begin{figure}[!tp]
    \centering

    \begin{adjustbox}{max width=\textwidth, max height=0.725\textheight}
        \includegraphics{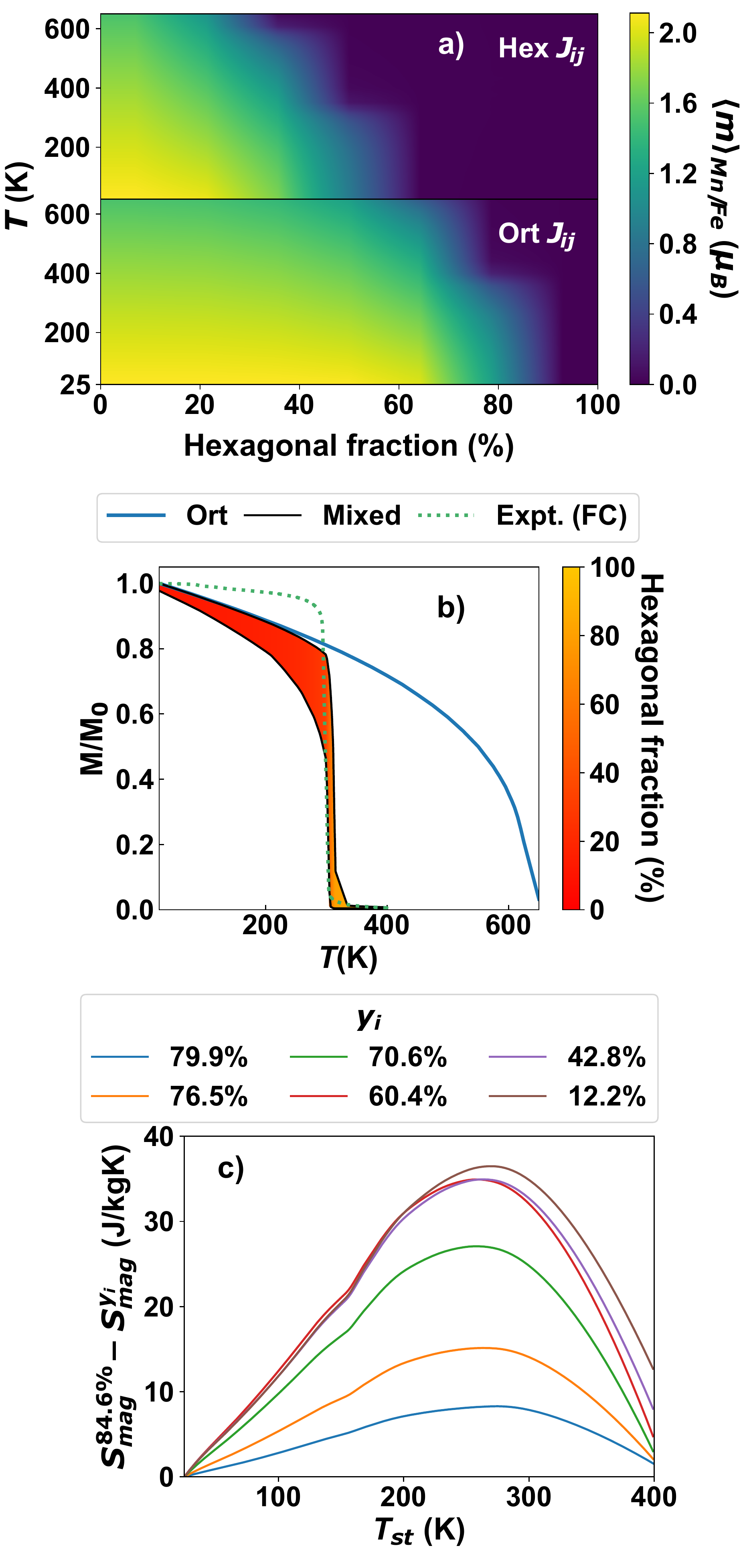}
     \end{adjustbox}
  
        \caption{(a) Calculated  average magnetization as a  function of temperature and the hexagonal phase weight-$\%$ for the Mn$_{0.5}$Fe$_{0.5}$NiSi$_{0.95}$Al$_{0.05}$ compound.  Calculated (solid lines) and measured (green dotted line) temperature dependent normalized magnetization. The results from MC simulations for the pure orthorhombic phase and the mixed magnetic composite phase are plotted in blue and black solid lines, respectively. The upper and lower curves of the mixed phase refer to the two extreme cases for the unknown $J_{ij}^{inter}$ and define the region where the true result for the composite phase is expected. The region has been coloured according to the temperature dependent hexagonal weight-$\%$ derived from the XRPD results (excluding the impurity phase). (c) Calculated change of magnetic entropy associated with a magnetic field-induced increase of the hexagonal phase fraction from $y_i$ to $y_f$ at the structural transition temperature $T_{st}$. The different curves correspond to different initial hexagonal weights $y_i$. The value 84.6\% corresponds to the maximum value of $y$ for the $x = 0.5$ compound at $T > T_{st}$.}
    \label{fig:mixing_M} 

\end{figure}

From the previous discussion, it is clear that the magnetic and structural transitions are coupled in the MnFeNiSiAl compounds, and the coexistence of the two phases in the experimental samples is important to take into account for realistic simulations of the magnetic properties. 
A model describing this situation has been proposed by Skomski and Sellmyer \cite{skomski2000curie}, where the magnetic ordering temperature was calculated for a nanocomposite material consisting of two coexisting magnetic phases, one with low and one with high magnetic transition temperature. The calculations of Ref. \cite{skomski2000curie} showed that there will only be one magnetic transition temperature, even for a composite phase with distinctly different Heisenberg interactions. As discussed in this work, the $T_C$ of the composite phase, in general, was found to be larger than the volume average of the transition temperatures for the involved phases. 

The situation for the MST of MnFeNiSiAl is more complex than the model described in Ref. \cite{skomski2000curie}, since the weight percent of the involved phases is strongly temperature dependent in the transition region (cf. Fig. \ref{XRD_MCE}(b)). To explore the idea of the magnetic composite the calculated exchange interactions within a cutoff radius of $d/a=2.5$ of the orthorhombic phase were mapped onto the hexagonal phase for the Mn$_{0.5}$Fe$_{0.5}$NiSi$_{0.95}$Al$_{0.05}$ compound. Then for the atomistic spin simulations, the magnetic composite was treated as a disordered alloy with a mixture of parameters from the hexagonal phase, $y$, and the orthorhombic phase, $(1-y)$.
Therefore, there are in principle four types of atoms included in the atomistic simulation box: the hexagonal Mn$_{hex}$  and Fe$_{hex}$ atoms that amount individually to a fraction occupation of $y/2$, and the orthorhombic Mn$_{ort}$ and Fe$_{ort}$ atoms that amount individually to $(1-y)/2$. Note that the interface exchange interactions between species of different phases ($J_{ij}^{inter}$) are unknown and their calculation is non-trivial. However, it is reasonable to assume that the $J_{ij}^{inter}$ parameters would be somewhere in between the set of  $J_{ij}$ from the orthorhombic phase $J_{ij}^{ort}$ and the ones of the hexagonal phase $J_{ij}^{hex}$. Hence, the two extreme cases $J_{ij}^{inter} = J_{ij}^{ort}$ and $J_{ij}^{inter} = J_{ij}^{hex}$, have been considered to simulate the behaviour of the magnetic composite phase, such that a window of the expected magnetic properties can be established.

Figure~\ref{fig:mixing_TC} shows the computed magnetic ordering temperature for the magnetic composite as a function of the weight percent of the hexagonal phase. It can be seen that the window for a room temperature $T_C$ ranges from 40\% concentration of hexagonal phase to a concentration of 75\%. In addition to the Curie temperature also the average magnetization (cf. Fig. \ref{fig:mixing_M}(a), with details given in SI) and the magnetic structure are affected by the mixture of orthorhombic and hexagonal phases, as can be seen in the upper and bottom panels of Fig.~\ref{fig:mixing_TC}. 
Using $J_{ij}^{ort}$ for the interface exchange constants conserves  ferromagnetic ordering at low fractions of the hexagonal phase yielding a magnetic ordering temperature almost independent of  the fraction of the hexagonal phase (cf. Fig.~\ref{fig:mixing_M}(a)). Assuming instead that the interface exchange interactions are dominated by  $J_{ij}^{hex}$ leads to a fast drop of the ferromagnetic ordering temperature with  an increasing amount of the hexagonal phase. In a realistic scenario, the exchange interactions will be a mixture of both extremes such that with an increasing fraction of the hexagonal phase the 
 competition between the AFM couplings from the hexagonal phase and the FM couplings from both phases will also increase, which in order to be accommodated favour an increasingly non-colinear arrangement of the magnetic moments.


These results show that the coexistence of phases can have a considerable impact on the magnetic properties and suggest that the structural component is driving the MST, since the growth of the hexagonal phase in the magnetic composite drives $T_C$ to the experimental range of magnetic transition temperatures. In order to explore this idea, our model has been expanded to include the variation of the hexagonal phase with temperature. From the calculated temperature dependent magnetization curves, a magnetization surface $M(T,y)$, being a function of temperature ($T$) and the hexagonal phase fraction in the composite ($y$), has been constructed (cf. Fig. \ref{fig:mixing_M}(a)). Combining this data with the hexagonal phase weight-$\%$ derived from the XRPD results  (cf. Fig.~\ref{XRD_MCE}(b)) allowed us to calculate the expected temperature dependence of the magnetization taking into account the temperature dependence of the hexagonal phase.  
The results are displayed in Fig.~\ref{fig:mixing_M}(b) in which the experimental data has been included for comparison as well as the magnetization curve simulated for the pure orthorhombic phase. The proposed model provides a good description of the experimental results, supporting the picture of an MST  with the loss of magnetization and magnetic order induced by the structural change. The fact that the structural component is the leading element of the MST, explains the weak influence of external magnetic fields on $T_C$ observed in the experiments, (\textcolor{green}{cf. SI.\ref{SI:mixing} for details}). Moreover, a $T_C$ of $455$ K has previously been reported for the Mn$_{0.5}$Fe$_{0.5}$NiSi compound \cite{biswas2019designed}, considerably higher than the one measured here for  Mn$_{0.5}$Fe$_{0.5}$NiSi$_{0.95}$Al$_{0.05}$ ($\approx 300$ K). Such small doping with Al should not alter the magnetic properties of the pure hexagonal and orthorhombic phases. Instead, the difference in $T_C$ values can be linked to the lower concentration of the hexagonal phase observed at low temperature in Ref. \cite{biswas2019designed}, supporting the hypothesis that it is the mixture of hexagonal and orthorhombic phases which determines the MST and in particular the magnetic transition temperature of the composite material.  

Another argument for the structural transition leading the MST can be found from the rate of decrease of the magnetic ordering temperature with increasing concentration of Fe. A comparison between the theoretically calculated $T_C$ values for the pure phases and the experimental values in Fig.~\ref{Theory}(b), reveals that theory predicts a more modest decrease rate than the one measured. Note that the calculations only consider the magnetic subsystem, i.e.  the calculations do not consider the weights of the hexagonal and orthorhombic phases and how these are affected by the Fe concentration. The phase weights for the $x=0.50-0.55$ compounds are shown in Fig.~\ref{XRD_MCE} (b). At  constant temperature below the structural transition temperature $T_{st}$, the phase weight for the hexagonal phase increases with increasing $x$.  Therefore, it is clear that the concentration of Fe helps to stabilize the hexagonal phase.

The heat capacity of the spin system ($C_{mag}$) was calculated to estimate the magnetic entropy contribution to the GMCE. The calculations were performed for zero magnetic field and a magnetic field of 
$H = 2$T, with $J_{ij}^{inter} = J_{ij}^{ort}$ ({\color{green} cf. SI.\ref{SI:mixing} for details)}. The results reveal that close to room temperature the temperature dependence of the magnetic energy is dominated by the temperature dependence of the hexagonal phase suppressing the effect of  
the magnetic field. Thus, calculations using the measured temperature dependent weights for the hexagonal phase ($y(T)$) of the $x = 0.5$ compound, as used for the calculation of the magnetization in Fig.~\ref{fig:mixing_M} (b), would underestimate the change of magnetic entropy since $y(T)$ under a magnetic field is unknown. Instead, the change of magnetic entropy is calculated as:

\begin{equation}
    \small
    S_{mag}^{y_{f}}(T_{st})-S_{mag}^{y_{i}}(T_{st})=\int_0^{T_{st}} \frac{C_{mag}(T,y_{f})-C_{mag}(T,y_{i})}{T} dT. 
    \label{eq:mag_entrpy}
\end{equation}
In this expression, we assume a sharp first-order structural transition and take into account that $\Delta S_{mag}= S_{mag}^{y_{f}} - S_{mag}^{y_{i}}$ for a magnetic composite will depend on the initial ($y_{i}$) and final ($y_{f}$) values of the hexagonal phase concentration. 
While the magnetic field dependence is not explicit in Eq. (\ref{eq:mag_entrpy}), the difference between $y_{i}$ and $y_{f}$ can be interpreted as the change of the hexagonal phase weight induced by a magnetic field.

From Fig.~\ref{XRD_MCE} (b) it is clear that at high temperature ($>T_{st}$), the maximum value of $y$ is $\approx84.6\%$ for the  $x=0.5$ compound, which we take as $y_{f}$ in the calculations. However, the value of $\Delta S_{mag}$ will be different if $y_{f}$ is reached from different initial values of $y_{i}$ near the beginning of the structural phase transition, as shown in Fig.~\ref{fig:mixing_M}(c). 
It can be noted that the magnitude of the calculated $\Delta S_{mag}$ is larger than the measured value of 20 J/(kg K) for  $y_{i}\leq$70\%, but the correct order of magnitude of this calculation highlights the importance of the magnetic field induced change of the hexagonal phase concentration to explain the measured values of the isothermal entropy change. Besides, the lattice contribution to the entropy variation often counteracts the magnetic contribution, reducing the total value of the entropy change. The calculated $\Delta S_{mag}$ also highlights that a relatively small field induced change of the hexagonal concentration is enough to achieve a significant entropy change.
Moreover, the peak value of $\Delta S_{mag}$ close to the measured transition temperature suggests that the magnetic entropy also plays a role in triggering the structural transition. Further verification of this coupling requires the computation of the free energies. However, the computation of the vibrational entropy constitutes a problem as the Debye model is not reliable in these compounds, also observed for FeNiSi in Ref.~\cite{Fortunato2023}, and the hexagonal phase is dynamically unstable at $0$ K, not allowing a standard treatment within the harmonic approximation.

\subsection*{Conclusions}

In conclusion, we have shown that it is the mixture and coexistence of the hexagonal and orthorhombic phases (that have a dichotomy in the nature of the interatomic exchange) in the Mn$_{1-x}$Fe$_x$NiSi$_{0.95}$Al$_{0.05}$ system that determines the properties of the MST. Considering the experimentally determined temperature dependence of the two phases close to the structural transition $T_{st}$ and using the interatomic exchange parameters for the two phases calculated by ab-initio theory as input in atomistic Monte Carlo simulations, it is shown that the magnetic ordering temperature $T_C$ of the composite material can be quantitatively predicted. Experimental results reveal that 
the strength of the MST of the Mn$_{1-x}$Fe$_x$NiSi$_{0.95}$Al$_{0.05}$ system, and its relevance for the GMCE, is governed by the sharpness of the structural transition with respect to temperature. Moreover, theoretical calculations of the magnetic entropy contribution to the GMCE highlights the importance of a magnetic field induced change of the phase concentrations at the MST to explain the measured isothermal entropy change. This finding calls for in-field experimental studies of the structure close to the MST.
Overall we argue that the approach used here combining ab-initio theory with experimental studies with a focus on the mixture and coexistence of structural phases close to a MST, provides  
a new avenue to find materials with improved GMCE and forms a pathway for further investigations. 


\subsection*{Experimental and theoretical methodology}
All compounds studied in this work were synthesized using an arc-melting process. Pieces of the high-purity metals Mn, Fe, Ni, Si, and Al were weighted in certain proportions. During arc-melting, oxygen contamination was minimized by flushing the furnace five times with Ar and by melting a Ti getter prior to melting the sample. The samples were turned over and re-melted three times to promote homogeneity. Negligible losses of Mn ($<1$ weight-\%) were detected in most cases. In the event of larger loss, more Mn was added to compensate as only Mn evaporates in these alloys during arc-melting. Further, samples were wrapped in Ta-foil, placed in a quartz tube, and sealed under vacuum for later annealing at $1073$ K for $1$ week. Afterward, samples were quenched in cold water. X-ray powder diffraction (XRPD) data were  collected at different temperatures ranging from $265$ K to $422$ K using a Bruker D$8$ Advance diffractometer with Cu-K$_{\alpha}$ radiation, with an angle step size of 0.02${^\circ}$. The microstructure was investigated using a Zeiss Merlin scanning electron microscope (SEM) equipped with a secondary electron (SE) detector and an energy-dispersive X-ray spectrometer (EDS). Thermal analysis in the temperature range  from $190$ K to $650$ K was performed using a $204F1$ Netzsch Heat-Flux differential scanning calorimeter with aluminium pans. The magnetic properties were investigated in the temperature range from $5$ K to $400$ K using the Magnetic Property Measurement System (MPMS) and Physical Property Measurement System  (PPMS) from Quantum Design. 

Magnetic moments and magnetic exchange interactions ($J_{ij}$) were calculated  for the experimental structures and compositions (atomic occupancy and lattice parameters) with the spin-polarized Korringa-Kohn-Rostoker method~\cite{korringa1947calculation,kohn1954solution} within the atomic sphere approximation (ASA) as implemented in the SPR-KKR code~\cite{Ebert2011CalculatingApplications}. The chemical disorder was treated within the coherent potential approximation (CPA) \cite{Soven1967, cpa2}. The one-electron equations were solved within the scalar-relativistic approximation with a minimal basis set ($s$,$p$ and $d$ states). The calculations were performed  using the Perdew-Wang \cite{Perdew1992AccurateEnergy} formulation of the local spin density approximation for the exchange-correlation functional. For the determination of the $J_{ij}$ parameters  the Lichtenstein–Katsnelson–Antropov–Gubanov formalism~\cite{Liechtenstein1987}  was considered for the calculation of the Mn and Fe couplings within a cutoff radius of 2.5 (5.0 on the hexagonal cells) lattice units. Including Ni exchange couplings do not significantly alter the mean-field magnetic ordering temperatures for the orthorhombic and hexagonal phases. Including the Ni exchange couplings yield 833K and 339K for the magnetic ordering temperatures of the orthorhombic and hexagonal phases, respectively, which can be compared to the corresponding values of 828K and 338K excluding the same exchange parameters, justifying our simplification. The magnetic moments and the exchange parameters $J_{ij}$ derived from first-principles calculations were used as input for Monte Carlo (MC) simulations. 

The MC simulations were executed with the Uppsala atomistic spin dynamics (UppASD) code \cite{ASD1, ASD2} using the atomistic Heisenberg Hamiltonian: 
\begin{equation}
    \mathcal{H} = -{1 \over 2} \sum_{i \neq j} J_{ij} \mathbf{e_i} \cdot \mathbf{e_j},
    \label{mag_H}
\end{equation}
which describes the pair exchange interactions  between normalized magnetic moment vectors (\textbf{e}). In these simulations, a simulation box of $24\times 24\times 24$ unit cells (55296 spins) with periodic boundary conditions was considered. For each temperature step, we allow for 50000 MC steps for equilibration followed by $10^5$ steps for measurement of the  thermodynamic averages. From the average energy (defined by Eq.~\ref{mag_H}) for the different temperatures ($T$), the heat capacity was calculated as $\partial \langle E \rangle / \partial T$ to determine the magnetic ordering temperature. For improved accuracy, a temperature step of 5K was considered around the peak region. Since the heat capacity peaks at phase transitions, it is a good indicator to find $T_C$, having the advantage over the magnetisation analysis that it allows identifying other order-disorder magnetic phase transitions. This choice was motivated by the  results of the Monte Carlo simulations for the hexagonal phase at low temperatures, in which case  a non-collinear magnetic configuration is observed  .

\subsection*{Acknowledgments}
The authors thank the Swedish Foundation for Strategic Research (SSF), project "Magnetic materials for green energy technology” (contract EM$-16-0039$) for financial support. The authors acknowledge support from STandUPP and eSSENCE. O.E. acknowledges support from the Knut and Alice Wallenberg Foundation, the Swedish Research Council (VR) and the ERC (FASTCORR project). RMV thanks Nuno Fortunato for the discussion related to the Debye model calculations. S.I.S. acknowledges the support from the Swedish Research Council (VR) (2019-05551), the Swedish Government Strategic Research Areas in Materials Science on Functional Materials at Link\"{o}ping University (Faculty Grant SFO-Mat-LiU No. 2009-00971). The computations were enabled by resources provided by the National Academic Infrastructure for Supercomputing in Sweden (NAISS) and the Swedish National Infrastructure for Computing (SNIC), partially funded by the Swedish Research Council through Grant Agreements No. 2022-06725 and No. 2018-05973, and by the CSC – IT Center for Science, Finland.   

\subsection*{Author contributions}
SG and RMV took the leading role in the experimental and theoretical analysis of this work, respectively, as well as prepared the first draft of the manuscript. VS prepared the compounds and performed the structural analysis. MS and PS supervised the experimental parts of this work. SIS constructed the special quasi-random, ordered, and phase-separated supercell structures. OE proposed to explore the phase mixing of the composite material. EKD-C and RMV performed the first-principles calculations and data analysis. TB and HCH provided technical support for the DFT calculations. HCH,  OE and TB supervised the theoretical parts of this work. All authors discussed the results and contributed to writing of the manuscript.  
\subsection*{Competing interests}
Te authors declare no competing interests.

\subsection*{Data availability}
All data that support the findings of this study are included within the article (and any supplementary files).

\bibliography{ref.bib}
\bibliographystyle{prb-titles.bst}

\clearpage

\begin{widetext}
\section*{Supplementary Information (SI)}
\subsection{Effect of impurity phase}
\label{SI:impurity}
An MgZn$_2$-type hexagonal impurity phase has been observed for all studied compounds as seen in Fig. \ref{SEM_xrd}(a) for the XRPD refinement of the $x = 0.5$ compound. A likely similar (reported as unknown) impurity phase was observed by Biswas \emph{et al.} \cite{biswas2019designed}. From the SEM and EDS analysis, the chemical compositions of different surface regions (denoted as 1,2 and 3 in Fig.\ref{SEM_xrd}(b)) for the $x=0.5$ compound have been determined. Regions 1 and 2 have a chemical composition Mn$_{0.46}$Fe$_{0.55}$Ni$_{1.1}$Si$_{0.85}$Al$_{0.05}$ and Mn$_{0.46}$Fe$_{0.49}$Ni$_{1.16}$Si$_{0.85}$Al$_{0.05}$, respectively which is as expected within the error limit of the EDX analysis. However, region $3$ exhibits a Fe-deficient chemical composition,   Mn$_{0.9}$Fe$_{0.1}$Ni$_{1.2}$Si$_{0.7}$Al$_{0.04}$, and can possibly correspond to the MgZn$_2$-type hexagonal impurity phase observed in XRPD.

\begin{figure}[ht]
    \centering
    \includegraphics[width=\linewidth]{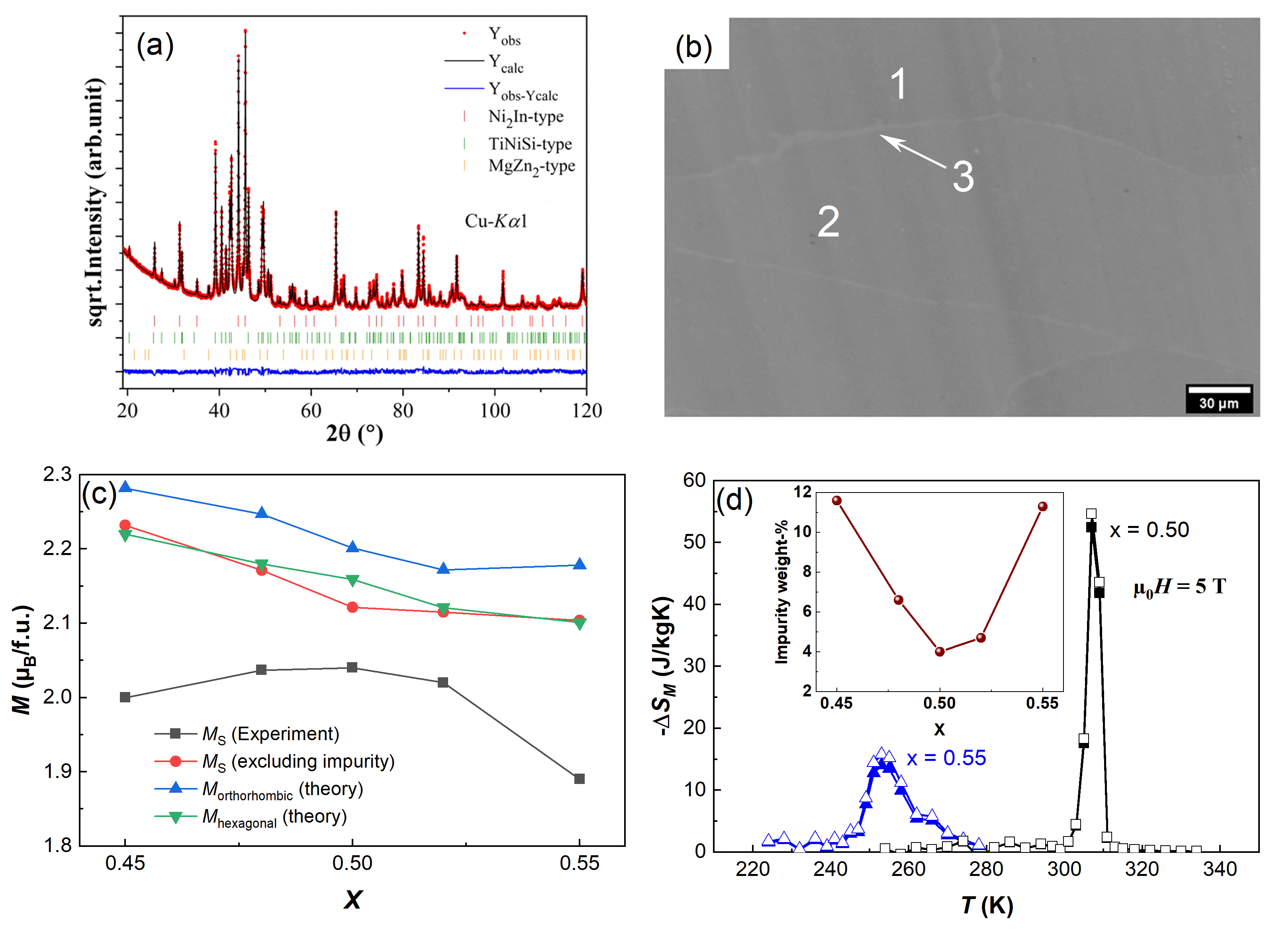}
    \caption{(a) XRPD refinement for the $x=0.5$ compound. The  data was collected at ambient conditions. (b) SEM/EDS analysis of the $x=0.5$ compound, the white regions correspond to the impurity phase, which has the chemical composition Mn$_{0.9}$Fe$_{0.1}$Ni$_{1.2}$Si$_{0.7}$Al$_{0.04}$. (c) Calculated  magnetic moment per formula unit and the measured saturation magnetization of the studied compounds. (d) Isothermal entropy change with (solid symbols) and without (hollow symbols) considering the impurity phase. Inset shows the weight-\% of the impurity phase.} 
    \label{SEM_xrd} 
\end{figure}

The weight-\% of the impurity phase  increases if the Fe/Mn ratio deviates from unity, as is seen in the inset of Fig. \ref{SEM_xrd} (d). However, from magnetization measurements, no extra magnetic phase transition has been observed below $400$ K. The existence of a non-magnetic impurity phase will introduce  a systematic error when determining the magnetization of the compound. In Fig. \ref{SEM_xrd}(c), the measured values of the saturation magnetization ($M_S$) at $10$ K for the studied compounds are shown, which do not match with the theoretically calculated trend for the total magnetization. However, if the mass of the impurity phase is subtracted from the mass of the samples before calculating the magnetization of the compounds, the experimental values of $M_S(x)$ are in  agreement with the theoretical results. Similar mass corrections have been performed for the calculation of $\Delta S_M$ and the results are shown in Fig. \ref{SEM_xrd} (d). It can be concluded that the effect of this impurity phase is negligible for the calculated values of $\Delta S_M$.       

\subsection{Field dependence of $T_C$ and isothermal magnatization}
\label{SI:mag_expt}
From Fig.\ref{fig:Tc_field}(a) and its inset, it is clear that $T_C$ of the $x=0.5$ compound does not depend on the applied magnetic field. From the temperature dependent XRPD analysis it was observed that the structural transition in the $x=0.5$ compound occurs in the temperature range $300$ K to $315$ K. As a consequence of the MST, the isothermal magnetization curves exhibit a strong temperature dependence in this temperature range as shown in Fig. \ref{fig:Tc_field}(b).
\begin{figure}[ht]
    \centering
    \includegraphics[width=0.5\linewidth]{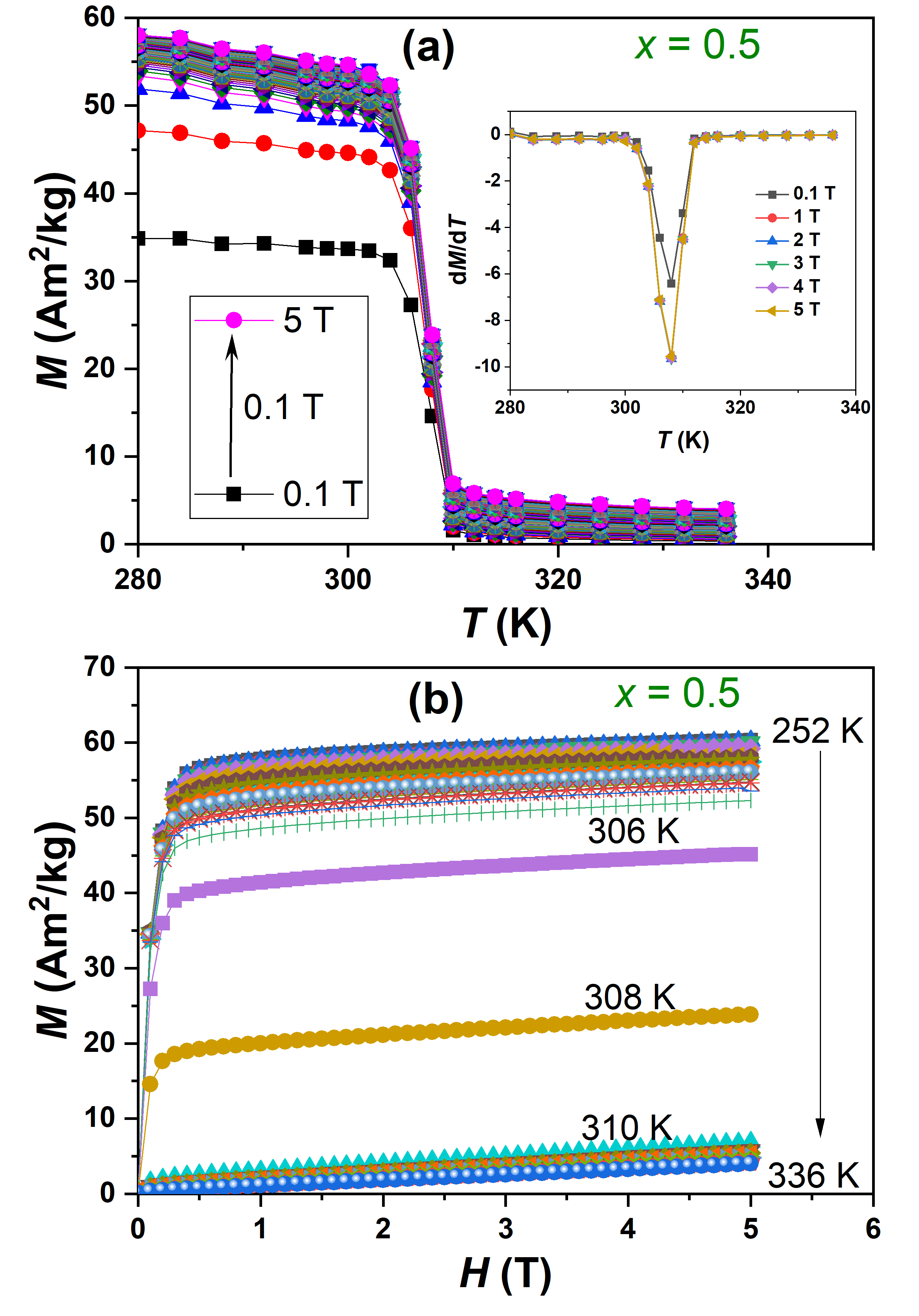}
    \caption{(a) Temperature dependent magnetization for different applied magnetic fields for the $x=0.5$ compound. The inset shows the first-order derivative of magnetization with respect to temperature. (b) Field dependent isothermal magnetization for the $x=0.5$ compound. All data shown in this figure were recorded following the phase reset cooling cycle protocol.} 
    \label{fig:Tc_field} 
\end{figure}

\subsection{Atomic site occupancy}
\label{SI:site_occ}
It is difficult to distinguish between Mn/Fe and Ni when determining which Wyckoff positions these species occupy using XRPD. From the literature, it is ambiguous if Ni occupies the $2a$ site or the $2d$ site in the high symmetry hexagonal phase, or if there is a random distribution of Mn/Fe/Ni between the $2a$ and $2d$ sites \cite{villars2022pearson}. To clarify this ambiguity, we compared the total electronic energy of the two possible arrangements, considering also intermediate site distributions with Ni mixing with Mn/Fe for the Mn$_{0.5}$Fe$_{0.5}$NiSi$_{0.95}$Al$_{0.05}$ composition. The total energies were calculated using the exact muffin-tin orbital (EMTO) method \cite{Andersen1994, Vitos2001, Vitos2007} for the different occupations with the soft-core approximation. As the calculations in the main text, the chemical disorder was treated within CPA \cite{Soven1967, Gyorffy1972} and the scalar-relativistic approximation using the Perdew-Wang \cite{Perdew1992AccurateEnergy}  exchange-correlation functional. The electrostatic correction to the single-site CPA was considered via the screened impurity model \cite{Korzhavyi1995} with a screening parameter of 0.6. The Green's function was calculated for 16 complex energy points distributed exponentially on a semi-circular contour including states within 1.1 Ry below the Fermi level. For the one-center expansion of the full charge density, a $l_{\rm max}^h$=8 cutoff was used.

The results in Fig.~\ref{fig:Theory_site} show that it is energetically favourable for Ni to occupy the $2d$ Wyckoff position, in agreement with Ref.~\cite{Fortunato2023}, and that it is reasonable to assume a small degree of mixing  ($<$20\%) between Ni and Mn/Fe sites for the annealing temperature used in the sample preparation. The calculations also predict that such an arrangement will maximize the total magnetic moment (see Fig.~\ref{fig:Theory_site}), mainly by the increase of the atomic magnetic moment of Ni (not shown).

\begin{figure}[ht]
    \centering
    \includegraphics [width=0.5\linewidth]{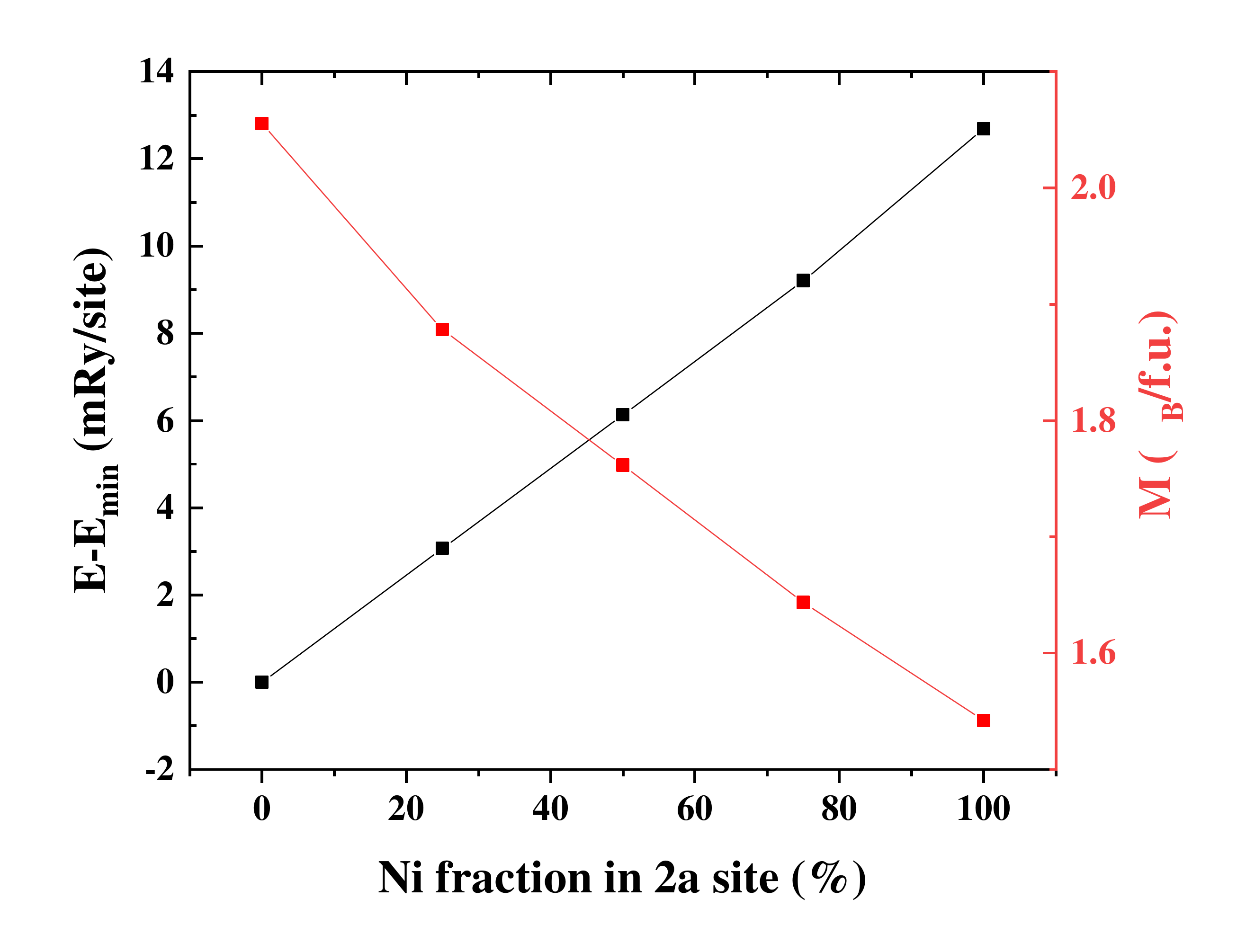}
    \caption{Calculated total electronic energy (black squares) and magnetic moment (red squares) versus Ni occupation of $2a$ Wyckoff positions in the hexagonal phase. Energies are displayed relative to the minimum energy.}
    \label{fig:Theory_site} 
\end{figure}


The high sensitivity on the Wyckoff positions and lattice parameters raises the question if one needs to consider thermal effects on the lattice parameters for an accurate  Curie temperature calculation. Still, the measured data (see Fig.~\ref{XRD_MCE}(c) ) reveal that the lattice parameter ratios of each phase do not vary significantly in the temperature range of the phase transition.

\subsection{Ni-Si bonds}
\label{SI:Ni_Si}

In Ref.~\cite{Landrum1998}, the structures of FeNiSi and MnNiSi in the orthorhombic phase were analyzed from neutron diffraction results and the alteration of the Ni-Si framework geometry if the interstitial space was populated either by Fe or Mn was discussed in detail by the authors. This different arrangement raises the question if in the  Mn$_{0.5}$Fe$_{0.5}$NiSi structure  the Ni-Si framework is arranged in an intermediate geometry or if it is dependent on the local environment. If the latter case is true, the description of the magnetic properties becomes more complex with the likelihood of local mechanisms. Since the analysis of the XRPD patterns cannot distinguish between these two cases, we performed structural relaxations in VASP ~\cite{Kresse1994,Kresse1996a,Kresse1996b} (PAW method) for 2$\times$2$\times$2 supercells with 3 different patterns of Mn-Fe occupation in Mn$_{0.5}$Fe$_{0.5}$NiSi (using as reference the structure determined for the Mn$_{0.5}$Fe$_{0.5}$NiSi$_{0.95}$Al$_{0.05}$ sample): one where there are two well-separated phases MnNiSi and FeNiSi; one where the occupation is mixed but ordered; one with random disorder (generated within the SQS method ~\cite{Zunger1990}). The calculations were performed with the PBE exchange-correlation functional (GGA)~\cite{PBE}, which generally performs well in the prediction of structural properties. Moreover, the standard PAW potentials provided by the code~\cite{Kresse1999} and a kinetic energy cutoff of 675 eV were considered. As convergence criteria for the structural relaxation, it was taken that the norms of all the forces should be smaller than 0.001 eV/\AA.

The Ni-Si framework relaxes homogeneously for all considered patterns, i.e.  without forming  localised features as can be seen in Fig.~\ref{fig:NiSi}. These results support that the structure refined from the XRPD pattern is not an average of local features but an actual representation of the crystalline structure. The lack of local geometry features also validates the use of the CPA to describe the chemical disorder in first-principles calculations.

\begin{figure}[!h]
    \centering
    \includegraphics[width=0.5\linewidth]{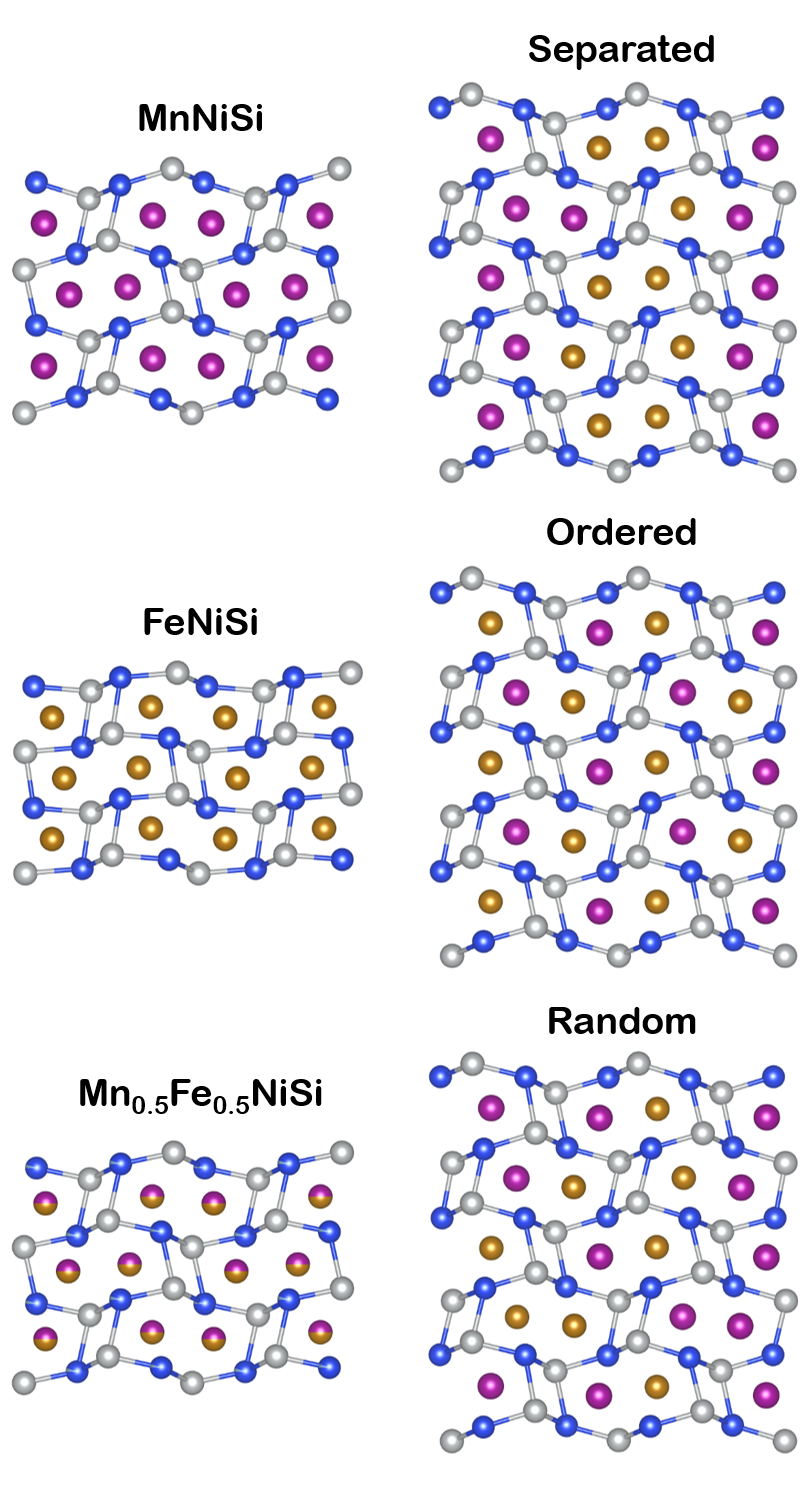}
    \caption{The Ni-Si chains have different geometries in the MnNiSi and FeNiSi compounds.  In the XRPD refined structure for the  Mn$_{0.5}$Fe$_{0.5}$NiSi$_{0.95}$Al$_{0.05}$ compound the Ni-Si bonds are arranged as an intermediate case between the parent compounds. Structural relaxations in supercells show that for different patterns such arrangement is indeed global, and not an average of local properties. The lack of local geometry features also corroborates the CPA approach taken. The colours distinguish the chemical species:  pink, blue, orange and grey stand for Mn, Si, Fe and Ni, respectively.}
    \label{fig:NiSi}
\end{figure}


\subsection{Spin Distribution}
\label{SI:spins_dist}
Figure ~\ref{fig:spins_dist} presents the statistical distribution of the relaxed spin orientations in the Monte Carlo simulations. In the case of the orthorhombic phase, the spins are oriented along a well-defined direction, more exactly along the $z$-direction since $\theta \approx 0^{\circ}$. Such a distribution corresponds to the expected ferromagnetic configuration. $\theta \approx 90^{\circ}$ corresponds to the case where the spins align in the $x-y$ plane. In the case of the hexagonal phase, one can distinguish three evenly spaced directions of the spin orientation in the $x-y$ plane. This three-fold symmetry may correspond to a frustrated spin configuration arising  from competing ferromagnetic and antiferromagnetic couplings, but further calculations are necessary to confirm such a hypothesis. With the data available, it is only possible to conclude that the hexagonal phase exhibits a non-collinear spin configuration with a finite magnetization due to the asymmetry in the spin distribution along the $z$-direction.

\begin{figure}[h]
    \includegraphics[width=0.3\linewidth]{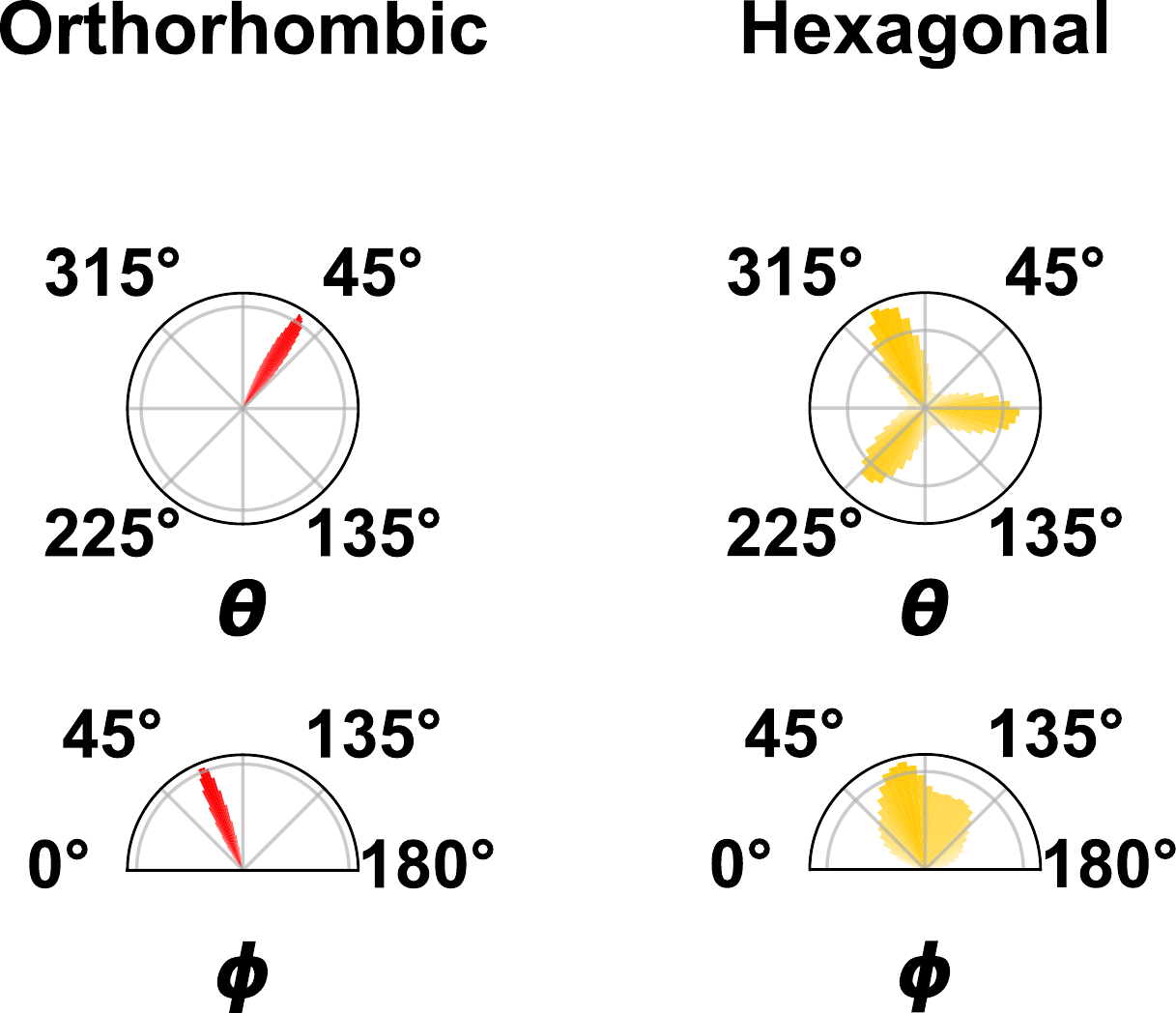}
    \caption{Statistical distribution of the spin orientations (spherical coordinates with polar angle $\theta$ and azimuthal angle $\phi$) relaxed in the magnetic simulations at $T=25$ K for the  orthorhombic (left) and hexagonal (right) phases.}
    \label{fig:spins_dist}
\end{figure}

\subsection{Effects of geometry alteration on the exchange parameters}
\label{SI:Jij_variation}


The structural similarity between the phases contrasts with the disparity in the respective magnetic properties.  In order to investigate the primary cause of this variation, two intermediate geometries were examined: an orthorhombic structure with lattice ratios equal to those in the hexagonal cell, and an orthorhombic structure with lattice parameters identical to those  in the hexagonal cell. The main results for the magnetic properties can be seen in Table~\ref{tab:mag_geometry}. There, from top to bottom, it is shown how atom displacement,  volume change and lattice ratio affect $T_C$.

It is observed that $T_C$ is not susceptible to volume change. However, lattice ratios and the atom's displacements strongly affect $T_C$. Analysis of the respective $J_{ij}$ (not shown) reveals that the Mn-Mn couplings around 5$\AA$ are responsible for the alteration, which are FM in the orthorhombic phase but become AFM in the hexagonal phase. 
 
\begin{table}[ht]
\centering
\caption{ Variation of calculated magnetic properties of  Mn$_{0.5}$Fe$_{0.5}$NiSi$_{0.95}$Al$_{0.05}$ with the geometrical parameters. The magnetic ordering temperature $T_C$ was calculated using the mean-field approximation. The experimental structures are highlighted in bold.}
\label{tab:mag_geometry}

\begin{tabular}{cccccc}
\toprule
\begin{tabular}[c]{@{}c@{}}Wyckoff \\ positions\end{tabular} &
 \begin{tabular}[c]{@{}c@{}}a \\ (\AA)\end{tabular} &
 b/a , c/a &
 \begin{tabular}[c]{@{}c@{}}Volume\\ (\AA$^3$)\end{tabular} &
 \begin{tabular}[c]{@{}c@{}}Mag. Mom. \\ ($\mu_B$/cell)\end{tabular} &
 \begin{tabular}[c]{@{}c@{}}T$_c$\\ (K)\end{tabular} \\ \hline
\textbf{hex phase} &\textbf{6.88} &\textbf{0.74, 0.58} & \textbf{139.53} & \textbf{8.63} & \textbf{305} \\
ort phase          & 6.88          & 0.74, 0.58          & 139.83          & 8.32          & 611          \\
ort phase          & 6.94          & 0.74, 0.58          & 143.54          & 8.58          & 617          \\
\textbf{ort phase} &\textbf{6.94} & \textbf{0.81, 0.53} & \textbf{142.79} & \textbf{8.80} & \textbf{823} \\

\bottomrule        
\end{tabular}%
\end{table}

\subsection{Orthorhombic-hexagonal mixing }
\label{SI:mixing}
In the Monte Carlo simulations for the magnetic composite, 7 different concentrations of hexagonal phase were considered: 0, 16.7, 33.3, 50, 66.7, 83.3, and 100 \%. As discussed in the main text, depending on the chosen set of $J_{ij}^{inter}$, the magnetic order is affected differently by the hexagonal fraction. In Fig.~\ref{fig:mixing_M0}, we observe for the $J_{ij}^{inter}=J_{ij}^{ort}$ scenario that with increasing hexagonal fraction  the magnetization remains constant until 66.7\%,  while in the other extreme, $J_{ij}^{inter}=J_{ij}^{hex}$ the magnetization drops rapidly with increasing hexagonal fraction. Note that from heat capacity calculations long-range magnetic ordering is observed for all composites,  
meaning that the drop in the magnetization is related to the increase of AFM couplings in $J_{ij}^{hex}$ (see Fig.~\ref{Theory}(c) in the main text) which compete with the FM couplings giving rise to a non-collinear arrangement of the spins.
\begin{figure}[!h]
    \centering
    \includegraphics[width=0.5\linewidth]{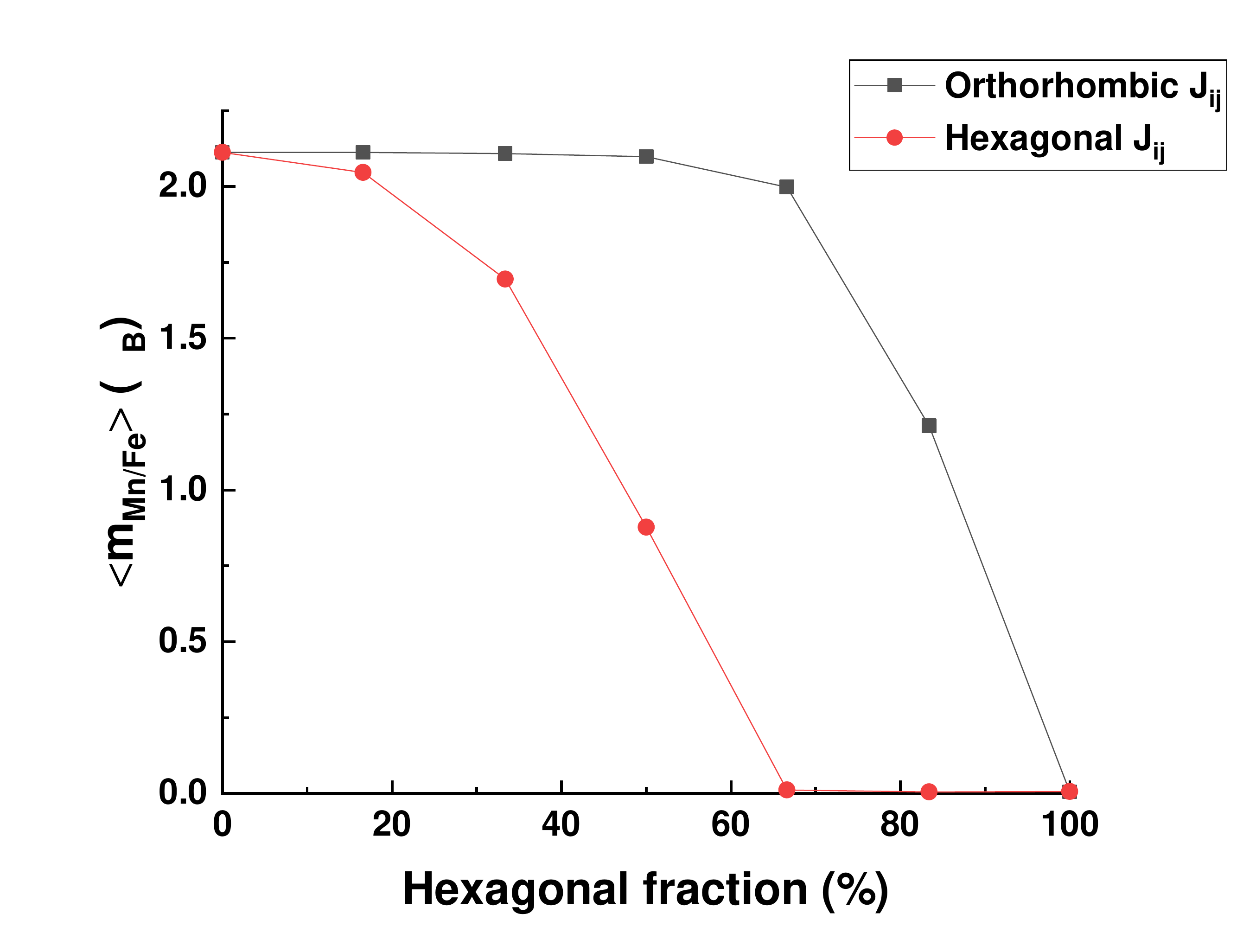}
    \caption{Calculated average magnetization for the composite phase at T=25K as a function of the hexagonal weight fraction. The data sets refer to two possible cases for the unknown interface exchange parameter $J_{ij}^{inter}$ between phases.}
    \label{fig:mixing_M0}
\end{figure}

The calculated temperature dependent magnetization for the different hexagonal fractions allows mapping a magnetization surface $M(T,y)$ (see Fig.~\ref{fig:mixing_M}(a) in the main text) being a function of temperature $T$ and the hexagonal fraction $y$. Applying a bilinear interpolation on $M(T,y)$ and combining with the hexagonal weights from the XRPD refinements (see Fig.~\ref{XRD_MCE} in the main text), renormalized to exclude the impurity phase, we can include the temperature dependent growth of the hexagonal phase in the simulation of the magnetization. The resultant temperature dependent magnetization using $J_{ij}^{inter}=J_{ij}^{ort}$, shown in Fig.~\ref{fig:mixing_M}(b) of the main text, drops in agreement with the experimental result close to $T_{st}$ capturing better the behaviour below $T_C$. Note that the experimental magnetization curve is stiffer at low temperatures than any of the calculated curves, being this a known consequence of a classical description of the spins in the simulations.

Given the better performance of the $J_{ij}^{inter}=J_{ij}^{ort}$ set, only this case was considered to study the magnetic field response. In Fig.~\ref{fig:M_curves} the temperature dependent magnetization curves can be seen for the different composites, with and without a magnetic file of 2 T. In general, for a given temperature, the change in magnetic response for a composite material exposed to a comparably large magnetic field is much smaller than the variation of magnetic response due to a change in material composition. Such behaviour is particularly evident at low temperatures but extends up to temperatures where the magnetic transition occurs. A consequence of this is that the magnetic field alone will not modify or change  the magnetic ordering, since this is ruled by the temperature dependent growth of the hexagonal phase close to the structural transition. However, the magnetic field should stabilize the high magnetization phase (the orthorhombic phase) and then change how the phases in the magnetic composite evolve with temperature. Thus, the magnetic field can indirectly affect the magnetic order and give rise to the measured GMCE.

\begin{figure}[!h]
   \centering
    \includegraphics[width=0.5\linewidth]{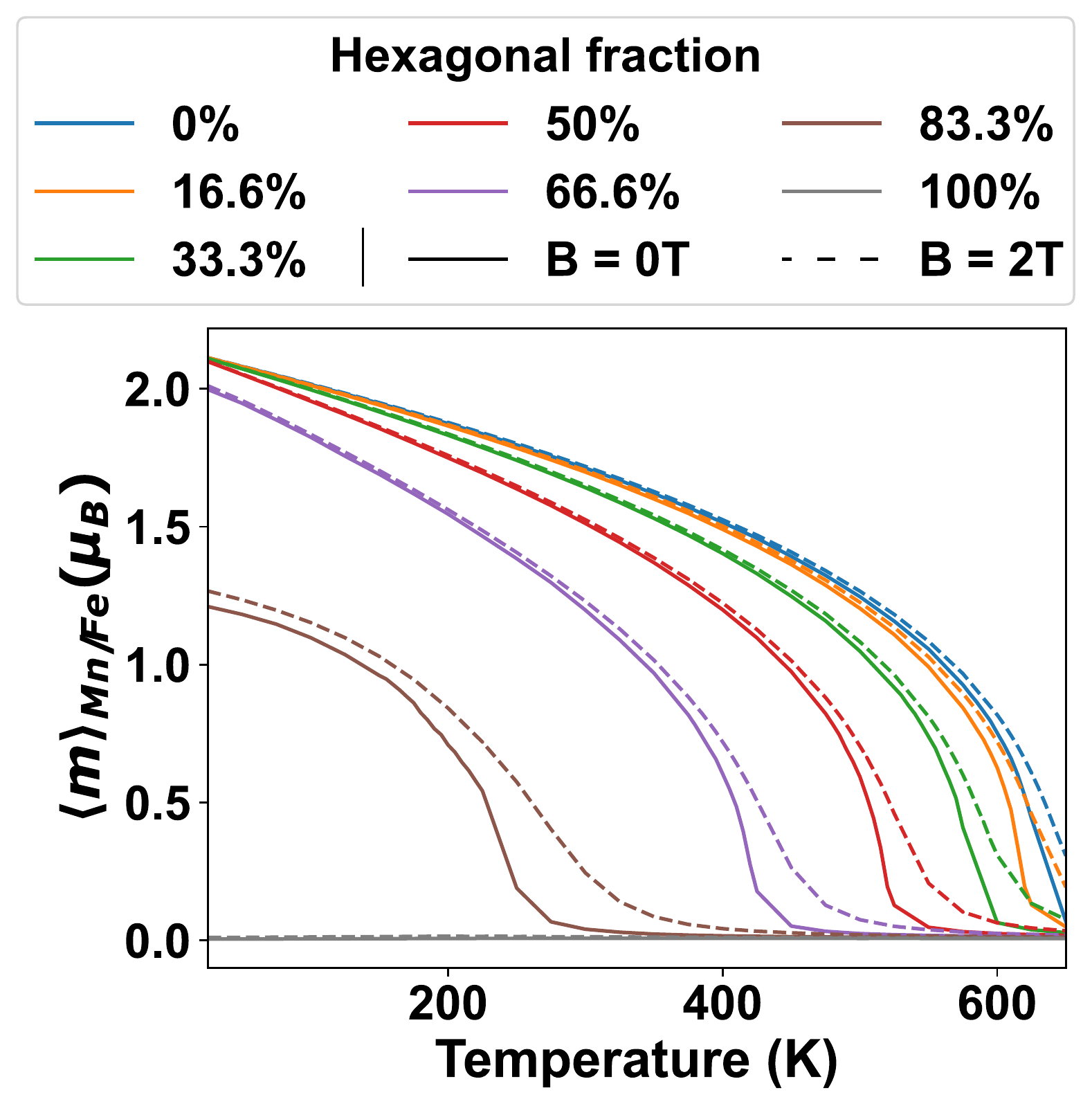}
   \caption{Temperature dependent magnetization curves from MC  simulations with (dashed lines) and without external magnetic field (solid lines). The colours distinguish different magnetic composites with distinct hexagonal weight fractions.}
   \label{fig:M_curves}
\end{figure}


\end{widetext}

\end{document}